\documentclass[review]{elsarticle}

\usepackage{lineno,hyperref}
\modulolinenumbers[5]

\journal{Journal of \LaTeX\ Templates}

\def\bea{\begin{eqnarray}}
\def\eea{\end{eqnarray}}

\begin{document}

\begin{frontmatter}

\title{Statistical-noise reduction in correlation analysis of high-energy nuclear collisions with event-mixing}

\author{R. L. Ray and P. Bhattarai}
\address{Department of Physics, The University of Texas at Austin, Austin, TX 78712 USA}




\begin{abstract}
The error propagation and statistical-noise reduction method of Reid and Trainor for two-point correlation applications in high-energy collisions is extended to include particle-pair references constructed by mixing two particles from all event-pair combinations within event subsets of arbitrary size. The Reid-Trainor method is also applied to other particle-pair mixing algorithms commonly used in correlation analysis of particle production from high-energy nuclear collisions. The statistical-noise reduction, inherent in the Reid-Trainor event-mixing procedure, is shown to occur for these other event-mixing algorithms as well. Monte Carlo simulation results are presented which verify the predicted degree of noise reduction. In each case the final errors are determined by the bin-wise particle-pair number, rather than by the bin-wise single-particle count.
\end{abstract}


\end{frontmatter}


\section{Introduction}
\label{Sec:Intro}
In correlation analysis of binned data the quantity of interest is the covariance of observable $x$ between arbitrary bins $m$ and $n$, given by
\begin{equation}
\overline{(x_m - \bar{x}_m)(x_n - \bar{x}_n)} = \overline{x_m x_n}
- \bar{x}_m \bar{x}_n,
\label{Eq1}
\end{equation}
where $x$ is the bin content and over-lines indicate averages over independent measurements. For example, in high-energy collisions between atomic nuclei $x$ could represent the number of sub-atomic particles produced and detected for each collision, or {\em event}, within bins defined by particle 3-momentum. For example, such bins could be constructed using the transverse momentum ($p_t$) (component of 3-momentum perpendicular to the direction of the colliding beams), the azimuth angle ($\phi$) in the plane transverse to the beam, the pseudorapidity ($\eta$) where $\eta = -\log(\tan{\theta/2})$ and $\theta$ is the polar angle relative to the beam direction, the azimuthal angle difference ($\phi_1 - \phi_2$) for arbitrary particles 1 and 2, and the pseudorapidity difference ($\eta_1 - \eta_2$). Quantity $x_m x_n$ is the number of particle-pairs in 2D bin $(m,n)$ and $\overline{x_m x_n}$ is calculated by averaging product $x_m x_n$ over all collision events in the event collection. Pairs of particles from the same event are referred to as {\em sibling} pairs. Averages $\bar{x}_m$ and $\bar{x}_n$ are calculated using all events in the collection where quantity $\bar{x}_m \bar{x}_n$ is the {\em reference}. For a given set of measured events the covariance will have a unique numerical value. The goal of this paper is to calculate the statistical error in this quantity by following and extending the error propagation and error reduction method of Reid and Trainor~\cite{TrainorReid}.

In practical applications measured quantities $x_m$, $x_n$ and $x_m x_n$ are affected by experimental inefficiency, acceptance and contamination. Inefficiency and acceptance losses in single particle counts and some contamination effects are readily corrected via ratio $\overline{x_m x_n}/(\bar{x}_m \bar{x}_n)$. However, two-particle inefficiencies, which occur when signals in the detectors from two particles are unresolved, cannot be corrected this way. Such inefficiencies are due to finite detector resolution and, if uncorrected, produce significant artifacts in the correlations for heavy-ion collisions~\cite{AxialCI,HBT}. The conventional correction method~\cite{HBT} involves removing particle pairs whose signals (e.g. induced ionization, secondary particle showers, etc.) fall within the resolution limits of the detector, and then removing pairs of particles from mixed-events which would have the same, relative locations in the detectors. Corrected results are obtained by increasing the minimum required separation between the detected signals of two nearby particles from zero until ratio $\overline{x_m x_n}/(\bar{x}_m \bar{x}_n)$ stabilizes. The practical consequence of this correction procedure is that the reference $ \bar{x}_m \bar{x}_n$ must be calculated by constructing uncorrelated pair counts in 2D bin $(m,n)$ by averaging over pairs of particles where each particle in a pair is selected from different collision events, referred to as {\em mixed-events}.

The statistical error of the covariance in Eq.~(\ref{Eq1}) equals the standard deviation of the distribution of covariance values corresponding to independent, statistically equivalent event samples (event collections) of underlying parent distributions for quantities $x_m$, $x_n$ and $x_m x_n$. Analytical calculations of this error therefore represent $x_m$, $x_n$ as random, event-wise fluctuating quantities relative to the parent distribution. The sibling and mixed-event pair-numbers are similarly represented. 

In Ref.~\cite{TrainorReid} Reid and Trainor derived a practical mixed-event method for calculating the reference in which random, event-wise fluctuations (noise) in $x_m$ and $x_n$, which are common to both sibling and mixed-event pair numbers, cancel in the covariance, significantly reducing the errors. For large data volumes summing the total number of mixed-event pairs can be computationally demanding. Various event-mixing algorithms have been developed by the heavy-ion community to reduce the necessary computation time while retaining sufficient statistical accuracy. One such method was discussed in Ref.~\cite{TrainorReid}. The choice of the reference and the event-mixing method strongly affects the statistical errors in the final correlation measurement. In this paper the statistical noise reduction method of Reid-Trainor will be extended and applied to other, practical event-mixing algorithms. The consequences of these event-mixing choices, or references, for the statistical uncertainties in the correlations will be quantified.

The present application is for ultra-relativistic heavy-ion collisions such as those measured by the STAR experiment~\cite{AxialCI,STAR} and by the experiments at the Large Hadron Collider (LHC)~\cite{LHC1,LHC2}. The methods presented here are also directly applicable to correlation analysis of multi-particle production from any type of particle collision. The present event-mixing technique for constructing an uncorrelated reference distribution has analogues in, for example, cosmology and acoustics. Measurements of the relative distance correlation between galaxies within an angular patch of the sky require an uncorrelated reference distribution. The latter can be constructed from cross-correlated pairs of galaxies observed in different sky patches or from randomly generated distributions of galaxies~\cite{Davis,Landy}. In acoustical analysis of multiple, independent time series the autocorrelation, or time-lag dependence, for each time series must be referenced to a cross correlation between two independent time series having the same lag time~\cite{acoustic}. In these two examples angular patches of sky or individual time series correspond to collision events and binned numbers of galaxies or acoustical amplitudes correspond to binned number of particles in the present analysis.

This paper is organized as follows. In Sec.~\ref{Sec:general} the Reid-Trainor procedure is derived and extended. In Sec.~\ref{Sec:other} their method is applied to other event-mixing algorithms. Monte Carlo studies are discussed in Sec.~\ref{Sec:montecarlo}. Conclusions are given in Sec.~\ref{Sec:conclusions}.

\section{Generalized Reid-Trainor event-mixing}
\label{Sec:general}

In Ref.~\cite{TrainorReid} the event collection was separated into pairs of events with similar, total number of detected particles, or {\em multiplicity}. Event-mixing was only applied between the two events in each pair. For arbitrary 2D bin $(m,n)$ ($m \neq n$) the total number of sibling pairs of particles in the collection is given by the sum
\bea
{\cal S}_{mn} & = & \sum_{g=1}^{N_g} \sum_{j=1}^2 (\bar{m}+\mu_j)(\bar{n} + \nu_j)
\label{Eq2}
\eea
and the mixed-event pair sum is given by
\bea
{\cal M}_{mn} & = & \sum_{g=1}^{N_g} \sum_{j^{\prime} > j=1}^2 
[(\bar{m}+\mu_j)(\bar{n} + \nu_{j^{\prime}}) + (\bar{m}+\mu_{j^{\prime}})(\bar{n} + \nu_j)]
\label{Eq3}
\eea
where indices $j$ and $g$ denote events and event groups, respectively. Variables $\bar{m}$ and $\bar{n}$ are the parent distribution number of particles in bins $m,n$ (bins are denoted with subscripts) and fluctuations are represented with random variables $\mu$ and $\nu$ as in Ref.\cite{TrainorReid}. Event group index $g$ is suppressed in the notation for fluctuations $\mu$ and $\nu$. The number of pairs of events in the collection is $N_g = N_{\rm events}/2$ where $N_{\rm events}$ is the number of events in the collection. Note that averages of random variables $\mu$ and $\nu$ within an event collection only vanish in the $N_{\rm events} \rightarrow \infty$ limit. For large event numbers the summations in Eqs.~(\ref{Eq2}) and (\ref{Eq3}) are approximately given by
\bea
{\cal S}_{mn} & \approx & N_{\rm events} (\bar{m}\bar{n} + \overline{\mu\nu})
\nonumber \\
{\cal M}_{mn} & \approx & N_{\rm events} (\bar{m}\bar{n}),
\label{Eq4}
\eea
where $\overline{\mu\nu} = (N_{\rm events}^{-1}) \sum_j \mu_j \nu_j$ is non-zero if the fluctuations in bins $m$ and $n$ are correlated. For relativistic heavy-ion collisions $\overline{\mu\nu}/(\bar{m}\bar{n}) << 1$~\cite{AxialCI,LHC1,LHC2} and for the purpose of calculating statistical errors the small contributions of $\overline{\mu\nu}$ can be neglected. The large event number limits are defined as $\bar{{\cal S}}_{mn} = \bar{{\cal M}}_{mn} = N_{\rm events} (\bar{m}\bar{n})$.

From the above discussion the correlation quantity of interest is ${\cal S}_{mn} - {\cal M}_{mn}$ and we calculate the number of correlated pairs per reference pair given by~\cite{AxialCI,LHC1,LHC2}
\bea
\frac{{\cal S}_{mn} - {\cal M}_{mn}}{{\cal M}_{mn}} & = & \frac{{\cal S}_{mn}}{{\cal M}_{mn}} - 1 \equiv {\cal R}_{mn} - 1.
\label{Eq5}
\eea
The statistical error in $({\cal S}_{mn} - {\cal M}_{mn})/ {\cal M}_{mn}$ equals the statistical error in ${\cal R}_{mn}$, denoted by $\Delta {\cal R}_{mn}$ and is given by
\bea
\left(\frac{\Delta {\cal R}_{mn}}{\bar{\cal R}_{mn}} \right)^2 & = &
\left(\frac{\Delta {{\cal S}}_{mn}}{\bar{{\cal S}}_{mn}} \right)^2   +
\left(\frac{\Delta {{\cal M}}_{mn}}{\bar{{\cal M}}_{mn}} \right)^2   -
\frac{ 2\Delta({\cal S},{\cal M})_{mn}}{\bar{{\cal S}}_{mn} \bar{{\cal M}}_{mn}}
\label{Eq6}
\eea
where $(\Delta {\cal R}_{mn})^2$, etc. are variances, $\Delta({\cal S},{\cal M})_{mn}$ is a covariance, and $\bar{\cal R}_{mn} = \bar{\cal S}_{mn}/\bar{\cal M}_{mn}$. Simplifying Eq.~(\ref{Eq6}) and averaging over event collections yields
\bea
(\Delta {\cal R}_{mn})^2 & = & \bar{\cal R}_{mn}^2
\frac{[\Delta({\cal S}_{mn} - {\cal M}_{mn})]^2}{\bar{{\cal M}}_{mn}^2}
\nonumber \\
 & = &
\frac{\langle [ ({\cal S}_{mn} - {\cal M}_{mn}) - \langle\bar{{\cal S}}_{mn} - \bar{{\cal M}}_{mn}\rangle ]^2 \rangle }
{\bar{{\cal M}}_{mn}^2} \nonumber \\
 & = & \frac{\langle ({\cal S}_{mn} - {\cal M}_{mn})^2 \rangle }{\bar{{\cal M}}_{mn}^2}
\label{Eq7}
\eea
where angle-brackets represent the average over independent event collections, $[\Delta({\cal S}_{mn} - {\cal M}_{mn})]^2$ is the variance of difference $({\cal S}_{mn} - {\cal M}_{mn})$, $\bar{\cal R}_{mn} = 1$, and $\langle\bar{{\cal S}}_{mn} - \bar{{\cal M}}_{mn}\rangle = 0$.

The key result of Ref.~\cite{TrainorReid} was to show that event-pair-wise mixing eliminates contributions of single-particle fluctuations, leaving only those contributions from fluctuations in the number of pairs. Using Eqs.~(\ref{Eq2}) and (\ref{Eq3}) the variance in the numerator of Eq.~(\ref{Eq7}) simplifies to
\bea
\langle ({\cal S}_{mn} - {\cal M}_{mn})^2 \rangle & = &
\langle [ \sum_{g=1}^{N_g} (\mu_1\nu_1 + \mu_2\nu_2 - \mu_1\nu_2 - \mu_2\nu_1) ]^2 \rangle \nonumber \\
 & = & \langle \sum_{j=1}^{N_{\rm events}} \mu_j^2 \nu_j^2 + \sum_{j=1,{\rm odd}}^{N_{\rm events}} (\mu_j^2 \nu_{j+1}^2 + \mu_{j+1}^2 \nu_j^2) \rangle
\label{Eq8}
\eea
where averages over products of bin-wise fluctuations from different events vanish (see Appendix A). Carrying out the above event averaging gives
\bea
\langle ({\cal S}_{mn} - {\cal M}_{mn})^2 \rangle & = &
N_{\rm events} \langle \sigma_{\mu}^2 \sigma_{\nu}^2 \rangle + N_g \langle \sigma_{\mu}^2 \sigma_{\nu}^2 + \sigma_{\mu}^2 \sigma_{\nu}^2 \rangle \nonumber \\
 & = & 2 N_{\rm events} \langle \sigma_{\mu}^2 \sigma_{\nu}^2 \rangle
\nonumber \\
 & \stackrel{\rm Poisson}{\longrightarrow} &
2 N_{\rm events} \langle \bar{m}\bar{n} \rangle \approx 2 N_{\rm events} \bar{m}\bar{n},
\label{Eq9}
\eea
where $\sigma_{\mu}^2$ and $\sigma_{\nu}^2$ are the variances in bins $m$ and $n$, respectively. These variances result from the following averages:
\bea
\frac{1}{N_{\rm events}} \sum_{j=1}^{N_{\rm events}} \mu_j^2 \nu_j^2
 & = & \sigma_{\mu}^2 \sigma_{\nu}^2, \nonumber \\
\frac{1}{N_{\rm events}} \sum_{j=1}^{N_{\rm events}} \mu_j^2 \nu_{j+1}^2 
& = & \sigma_{\mu}^2 \sigma_{\nu}^2. \nonumber 
\eea
These factorized results are valid if the single-particle fluctuations in bins $m$ and $n$ are uncorrelated, consistent with the above assumption that correlation corrections to the statistical errors are negligible. In the last two steps in Eq.~(\ref{Eq9}) the limit of Poisson distributions is assumed, denoted here and in the remainder of this paper with a limit symbol (arrow). In the last line in Eq.~(\ref{Eq9}) event collection average $\langle \bar{m}\bar{n} \rangle$ is approximated with $\bar{m}\bar{n}$ from a single event-collection.

Correlated single-particle fluctuations in the event-wise occupancies of different bins (other than that caused by collision dynamics) could also be significant if the range of multiplicities in the event collection is large or if the bin size is large relative to the detector acceptance. In such cases the event-wise frequency distributions of $\mu_j$ and $\nu_j$ would be inconsistent with a Poisson distribution. In Sec.~\ref{Sec:montecarlo} the frequency distributions for realistic bin sizes are shown to be accurately represented with Poisson distributions. Finally, the statistical error in ${\cal R}_{mn}$ for the event-mixing algorithm in Ref.~\cite{TrainorReid} is
\bea
\sqrt{(\Delta{\cal R}_{mn})^2} & = & \sqrt{\frac{2N_{\rm events}(\bar{m} \bar{n})}{(N_{\rm events}(\bar{m} \bar{n}))^2}} = \sqrt{\frac{2}{N_{\rm events}(\bar{m} \bar{n})}}.
\label{Eq10}
\eea
Note that this error equals $\sqrt{\bar{\cal S}_{mn}^{-1} + \bar{\cal M}_{mn}^{-1}}$, or the square-root of the sum of the relative variances of the sibling and mixed-event pair numbers and is not determined by the errors in the bin-wise single-particle number~\cite{TrainorReid}.

For 2D diagonal bins ($m=n$) only unique particle pairs are counted (no self-pairs) which requires a separate derivation. In this case ${\cal S}_{mn}$ and ${\cal M}_{mn}$ are given by
\bea
{\cal S}_{mm} & = & \sum_{j=1,{\rm odd}}^{N_{\rm events}} \left[ \frac{1}{2} (\bar{m}+\mu_j)
(\bar{m}+\mu_j - 1) \right. \nonumber \\
 &  & \left. + \frac{1}{2} (\bar{m}+\mu_{j+1})(\bar{m}+\mu_{j+1} - 1) \right]
\label{Eq11} \\
{\cal M}_{mm} & = & \sum_{j=1,{\rm odd}}^{N_{\rm events}} (\bar{m}+\mu_j) (\bar{m}+\mu_{j+1})
\label{Eq12}
\eea
and as before the large event number limits are
\bea
\bar{{\cal S}}_{mm} & = & \frac{1}{2} N_{\rm events} (\bar{m}^2 - \bar{m} + \sigma_{\mu}^2 )
\stackrel{\rm Poisson}{\longrightarrow} \frac{1}{2} N_{\rm events} \bar{m}^2
\label{Eq13} \\
\bar{{\cal M}}_{mm} & = &  \frac{1}{2} N_{\rm events} \bar{m}^2 .
\label{Eq14}
\eea
Inserting these expressions into the variance quantity yields
\bea
\langle ({\cal S}_{mm} - {\cal M}_{mm})^2 \rangle & = &
\frac{1}{4} \langle [ \sum_{j=1,{\rm odd}}^{N_{\rm events}}
( \mu_{j}^2 + \mu_{j+1}^2 -2 \mu_{j} \mu_{j+1} - 2\bar{m} - \mu_{j} - \mu_{j+1})]^2 \rangle
\nonumber \\
 & = & \frac{1}{4} \langle \sum_{j=1,{\rm odd}}^{N_{\rm events}}
(\mu_{j}^2 + \mu_{j+1}^2 -2 \mu_{j} \mu_{j+1} - 2\bar{m} - \mu_{j} - \mu_{j+1})^2 \rangle
\nonumber \\ 
 & = & \frac{1}{4} \langle \sum_{j=1}^{N_{\rm events}}
(\mu_{j}^4 + \mu_{j}^2 + 2\bar{m}^2 - 2\mu_{j}^3 - 4\bar{m}\mu_{j}^2) \nonumber \\
 & & + 6  \sum_{j=1,{\rm odd}}^{N_{\rm events}} \mu_{j}^2 \mu_{j+1}^2 \rangle
\label{Eq15}
\eea
where in going from the first to the second line in Eq.~(\ref{Eq15}) cross terms between different events cancel in the Poisson limit and in the last line we use the fact that event-collection averages of terms linear in the fluctuation vanish. Inserting the moments of Poisson distributions (see Appendix B), the resulting variance is
\bea
\langle ({\cal S}_{mm} - {\cal M}_{mm})^2 \rangle  
 & \stackrel{\rm Poisson}{\longrightarrow} &
N_{\rm events} \bar{m}^2
\label{Eq16}
\eea
and the statistical error in ${\cal R}_{mm}$, assuming Poisson distributions, is
\bea
\sqrt{(\Delta{\cal R}_{mm})^2} & = & \sqrt{\frac{N_{\rm events} \bar{m}^2}{(\frac{1}{2} N_{\rm events} \bar{m}^2)^2}} = \frac{2}{\bar{m}\sqrt{N_{\rm events}}}.
\label{Eq17}
\eea
The multiplicative factor in the numerator for this diagonal bin error is $\sqrt{2}$ larger than that for off-diagonal bins given in Eq.~(\ref{Eq10}) due to the reduced number of unique sibling and mixed-event pairs in diagonal bins.

In general the event collection could be sub-divided into an arbitrary number of groups of events with $\epsilon$ events per group where $1 < \epsilon \leq N_{\rm events}$ and with $\epsilon (\epsilon - 1)/2$ combinations of mixing event-pairs within each group. For this general case the sibling and mixed-event particle-pair sums and their limits are given by
\bea
{\cal S}_{mn} & = & \sum_{g=1}^{N_g} \sum_{j=1}^{\epsilon} (\bar{m}+\mu_j)(\bar{n}+\nu_j)
\nonumber \\
\bar{{\cal S}}_{mn} &  = &  N_g \epsilon \bar{m}\bar{n} = N_{\rm events} (\bar{m}\bar{n})
\label{Eq18} \\
{\cal M}_{mn} & = & \sum_{g=1}^{N_g} \sum_{j^{\prime} > j=1}^{\epsilon}
[ (\bar{m}+\mu_j)(\bar{n}+\nu_{j^{\prime}}) + (\bar{m}+\nu_{j^{\prime}})(\bar{n}+\mu_j) ]
\nonumber \\
\bar{\cal M}_{mn} &  = &   (\epsilon - 1) N_{\rm events} (\bar{m}\bar{n}).
\label{Eq19}
\eea
For $\epsilon > 2$ there are more mixed-event particle pairs than sibling and the ratio ${\cal R}_{mn}$ must be normalized such that the correlation signal, $({\cal S}_{mn} - {\cal M}_{mn})$, equals zero in the absence of correlations. The ratio ${\cal R}_{mn}$ for the general case is therefore defined as
\bea
{\cal R}_{mn} & = & \frac{\bar{{\cal M}}_{mn}}{\bar{{\cal S}}_{mn}}
 \frac{{\cal S}_{mn}}{{\cal M}_{mn}} = (\epsilon - 1)\frac{{\cal S}_{mn}}{{\cal M}_{mn}}
\label{Eq20}
\eea
and
\bea
{\cal R}_{mn} - 1 & = & \frac{(\epsilon - 1) {\cal S}_{mn} - {\cal M}_{mn}}{{\cal M}_{mn}}.
\label{Eq21}
\eea
The variance is evaluated as before where
\bea
\langle [ (\epsilon - 1) {\cal S}_{mn} - {\cal M}_{mn} ]^2 \rangle & = &
\langle \{ \sum_{g=1}^{N_g} [ (\epsilon - 1) \sum_{j=1}^{\epsilon} \mu_j \nu_j
- \sum_{j^{\prime} > j = 1}^{\epsilon} (\mu_j \nu_{j^{\prime}} +\mu_{j^{\prime}} \nu_j) ] \} ^2 \rangle
\nonumber \\
 & = & \langle \sum_{g=1}^{N_g} [ (\epsilon - 1) \sum_{j=1}^{\epsilon} \mu_j \nu_j
- \sum_{j^{\prime} > j = 1}^{\epsilon} (\mu_j \nu_{j^{\prime}} +\mu_{j^{\prime}} \nu_j) ]^2 \rangle
\nonumber \\
 & = & \langle \sum_{g=1}^{N_g} [ (\epsilon - 1) \sum_{j=1}^{\epsilon} \mu_j \nu_j
- \sum_{j^{\prime} \neq j} \mu_j \nu_{j^{\prime}} ]^2 \rangle
\label{Eq22}
\eea
where cross terms among different event groups vanish. This last expression can be further simplified using the fact that the event-collection average of $\mu_j \mu_{j^{\prime}}$ for different events $j^{\prime} \neq j$ vanishes, resulting in
\bea
\langle [ (\epsilon - 1) {\cal S}_{mn} - {\cal M}_{mn} ]^2 \rangle & = &
\langle \sum_{g=1}^{N_g} [ (\epsilon - 1)^2 \sum_{j=1}^{\epsilon} \mu_j^2
\nu_j^2 + \sum_{j^{\prime} \neq j=1}^{\epsilon} \mu_j^2 \nu_{j^{\prime}}^2
] \rangle
\nonumber \\
 &  & \hspace{-1.0in} = \langle \sum_{g=1}^{N_g} [ (\epsilon^2 - 2\epsilon +1 -1)
\sum_{j=1}^{\epsilon} \mu_j^2 \nu_j^2 + 
\sum_{j^{\prime}, j=1}^{\epsilon} \mu_j^2 \nu_{j^{\prime}}^2 ] \rangle
\nonumber \\
 &  & \hspace{-1.0in} =  N_g \epsilon \langle \frac{\epsilon^2 - 2\epsilon}{N_g \epsilon}
\sum_{g=1}^{N_g} \sum_{j=1}^{\epsilon} \mu_j^2 \nu_j^2
+ \frac{\epsilon^2}{N_g \epsilon} \sum_{g=1}^{N_g}
\frac{1}{\epsilon} \sum_{j=1}^{\epsilon} \mu_j^2
\frac{1}{\epsilon} \sum_{j^{\prime}=1}^{\epsilon} \nu_{j^{\prime}}^2 \rangle .
\label{Eq23}
\eea
The first term is an average over all events in the collection and results in the product of variances $\sigma_{\mu}^2 \sigma_{\nu}^2$ in the absence of correlated fluctuations between different bins. The second term is the average over all groups of the product of variances for each group. Eq.~(\ref{Eq23}) therefore reduces to
\bea
\langle [ (\epsilon - 1) {\cal S}_{mn} - {\cal M}_{mn} ]^2 \rangle & = &
N_g \epsilon \langle (\epsilon^2 - 2\epsilon) \sigma_{\mu}^2 \sigma_{\nu}^2
+ \frac{\epsilon}{N_g}  \sum_{g=1}^{N_g} (\sigma_{\mu}^2 + \delta_g)
(\sigma_{\nu}^2 + \zeta_g) \rangle
\nonumber \\
 &  & \hspace{-0.5in} = N_g \epsilon \langle \epsilon(\epsilon - 1) \sigma_{\mu}^2 \sigma_{\nu}^2 \rangle \stackrel{\rm Poisson}{\longrightarrow}  \epsilon(\epsilon - 1) N_{\rm events} (\bar{m}\bar{n}),
\label{Eq24}
\eea
where the variances within each group are expressed as the average variance for all event collections plus random fluctuations $\delta_g$ and $\zeta_g$, where averages $\langle \delta \rangle$, $\langle \zeta\rangle$, and $\langle \delta\zeta\rangle$ equal zero.

The error in an off-diagonal bin is therefore
\bea
\sqrt{(\Delta {\cal R}_{mn})^2} & = &
\sqrt{\frac{\epsilon(\epsilon - 1) N_{\rm events} \bar{m}\bar{n}}{((\epsilon - 1)N_{\rm events} \bar{m}\bar{n})^2}} = \sqrt{\frac{\epsilon/(\epsilon - 1)}{N_{\rm events} \bar{m}\bar{n}}}.
\label{Eq25}
\eea
This error equals that in Eq.~(\ref{Eq10}) when $\epsilon = 2$ and approaches $1/\sqrt{N_{\rm events} \bar{m}\bar{n}}$ as the number of events in each mixing group increases and approaches the entire number of events in the collection.  Again, note that this error equals $\sqrt{\bar{\cal S}_{mn}^{-1} + \bar{\cal M}_{mn}^{-1}}$. The improvement in statistical accuracy achieved by increasing $\epsilon$ can be weighed against the increased computational requirements in order to optimize the number of event groups to be used in the analysis of data.

\section{Other event mixing algorithms}
\label{Sec:other}
In typical analyses of heavy-ion collision data~\cite{MikeThesis,LizThesis} correlations are calculated by reading an unordered list of event data and sorting the particle kinematic information from those events into subsets corresponding to a global event-property such as overall multiplicity, a proxy for the overlap, or {\em centrality}, between the colliding nuclei. For each centrality all sibling particle-pairs are processed and to reduce computational requirements mixed-event particle-pairs are only processed for the first and second events in the collection, then for the second and third events, and so on until all events in the centrality bin have been processed. This event-mixing algorithm can be extended such that each event is mixed with the next two events in the list, or the next three events, etc. In this section statistical errors are derived for the above single, double and multiple-event mixing algorithms. It will be shown that the noise reduction obtained in Ref.~\cite{TrainorReid} also occurs for these other types of event-mixing methods.

\subsection{Single-event mixing method}
\label{Sec:single}
In this event mixing algorithm particle pairs are constructed for all events by mixing particles in event $j$ with those in event $j+1$. The number of sibling and mixed-event particle pairs and their limits are given by
\bea
{\cal S}_{mn} & = & \sum_{j=1}^{N_{\rm events}} (\bar{m}+\mu_j)(\bar{n}+\nu_j)
\equiv \sum_{j=1}^{N_{\rm events}} {\cal S}_{mn}^j \nonumber \\
\bar{\cal S}_{mn} & = & N_{\rm events} (\bar{m}\bar{n})
\label{Eq26} \\
{\cal M}_{mn} & = & \sum_{j=1}^{N_{\rm events} - 1} [(\bar{m}+\mu_j)(\bar{n}+\nu_{j+1})
+ (\bar{m}+\mu_{j+1})(\bar{n}+\nu_j)]
\nonumber \\
 & \equiv & \sum_{j=1}^{N_{\rm events} - 1} ({\cal M}_{mn}^{j,j+1} + {\cal M}_{mn}^{j+1,j})
\nonumber \\
\bar{\cal M}_{mn} & = & 2(N_{\rm events} - 1)\bar{m}\bar{n} \approx 2 N_{\rm events} (\bar{m}\bar{n}),
\label{Eq27}
\eea
where definitions ${\cal S}_{mn}^j$ and ${\cal M}_{mn}^{j,j^{\prime}}$ are introduced for brevity. The approximation $\bar{\cal M}_{mn} \approx 2 N_{\rm events} (\bar{m}\bar{n})$ occurs because the last event in the event list has no other event to mix with. For $N_{\rm events} >> 1$ those missing pairs can be ignored in the above summations.

In this single-event mixing algorithm there are approximately twice as many mixed-event pairs as sibling pairs. Therefore the variance to be calculated is given by
\bea
\langle (2{\cal S}_{mn} - {\cal M}_{mn})^2 \rangle & = &
\langle [ 2 \sum_{j=1}^{N_{\rm events}} {\cal S}_{mn}^j - 
\sum_{j=1}^{N_{\rm events} -1} ({\cal M}_{mn}^{j,j+1} + {\cal M}_{mn}^{j+1,j})]^2 \rangle.
\label{Eq28}
\eea
Twice the sum over sibling pairs can be written as two, identical summations. The resulting sibling and mixed-event sums can be separated into sums over even and odd values of event number $j$. As a result Eq.~(\ref{Eq28}) can be rearranged into sums with the identical structure as that in Ref.~\cite{TrainorReid}, thus achieving the same level of noise reduction for this alternate event-mixing algorithm.  The result is
\bea
\langle (2{\cal S}_{mn} - {\cal M}_{mn})^2 \rangle & = &
\langle [ \sum_{j=2,{\rm even}}^{N_{\rm events}} ({\cal S}_{mn}^j + {\cal S}_{mn}^{j+1} -{\cal M}_{mn}^{j,j+1} - {\cal M}_{mn}^{j+1,j}) 
\nonumber \\
 & + &  \sum_{j=1,{\rm odd}}^{N_{\rm events}} ({\cal S}_{mn}^j + {\cal S}_{mn}^{j+1} -{\cal M}_{mn}^{j,j+1} - {\cal M}_{mn}^{j+1,j}) ]^2 \rangle .
\label{Eq29}
\eea
From Ref.~\cite{TrainorReid} we find that
\bea
{\cal S}_{mn}^j + {\cal S}_{mn}^{j+1} -{\cal M}_{mn}^{j,j+1} - {\cal M}_{mn}^{j+1,j} & = &
\mu_j \nu_j + \mu_{j+1} \nu_{j+1} - \mu_j \nu_{j+1} - \mu_{j+1} \nu_j . \nonumber \\
\label{Eq30}
\eea
Substituting the above result into Eq.~(\ref{Eq29}) and combining the even and odd numbered summations gives
\bea
\langle (2{\cal S}_{mn} - {\cal M}_{mn})^2 \rangle & = &
\langle [ \sum_{j=1}^{N_{\rm events}} (2 \mu_j \nu_j - \mu_j \nu_{j+1} - \mu_{j+1} \nu_j)]^2 \rangle
\nonumber \\
 & = & \langle \sum_{j=1}^{N_{\rm events}} (2 \mu_j \nu_j - \mu_j \nu_{j+1} - \mu_{j+1} \nu_j)^2 \rangle
\label{Eq31}
\eea
where cross terms for different events vanish as before. Continuing, we find that the above equation in the Poisson limit reduces to
\bea
\langle \sum_{j=1}^{N_{\rm events}} (4 \mu_j^2 \nu_j^2 + \mu_j^2 \nu_{j+1}^2 + \mu_{j+1}^2 \nu_j^2) \rangle
 & = & 6 N_{\rm events} \langle \sigma_{\mu}^2 \sigma_{\nu}^2 \rangle
\nonumber \\
 &  \stackrel{\rm Poisson}{\longrightarrow} &  6 N_{\rm events} (\bar{m}\bar{n}).
\label{Eq32}
\eea

The statistical error for off-diagonal bins is
\bea
\sqrt{(\Delta {\cal R}_{mn})^2} & = & \sqrt{\frac{6 N_{\rm events}\bar{m}\bar{n}}{(2N_{\rm events}\bar{m}\bar{n})^2}} = \sqrt{\frac{3/2}{N_{\rm events}\bar{m}\bar{n}}}.
\label{Eq33}
\eea
This error is a factor of $\sqrt{3/4}$ smaller than that for the Reid-Trainor method due to the larger number of mixed-event pairs (twice as many pairs) generated by this algorithm. It is important to note that the noise reduction achieved via the mixing algorithm in Ref.~\cite{TrainorReid} is also achieved with the single-event mixing method and the error is determined by pair number where the error in Eq.~(\ref{Eq33}) equals $\sqrt{\bar{\cal S}_{mn}^{-1} + \bar{\cal M}_{mn}^{-1}}$.

For diagonal bins
\bea
{\cal S}_{mm} & = & \sum_{j=1}^{N_{\rm events}} \frac{1}{2}(\bar{m}+\mu_j)(\bar{m}+\mu_j - 1)
\nonumber \\
 & = & \sum_{j=1}^{N_{\rm events}} \frac{1}{2} ({\cal S}_{mm}^j -\bar{m} -\mu_j)
\nonumber \\
\bar{\cal S}_{mm} & = & \frac{1}{2} N_{\rm events} \bar{m}^2
\label{Eq34} \\
{\cal M}_{mm} & = & \sum_{j=1}^{N_{\rm events} - 1} (\bar{m}+\mu_j)(\bar{m}+\mu_{j+1}) = \sum_{j=1}^{N_{\rm events} - 1} {\cal M}_{mm}^{j,j+1}
\nonumber \\
\bar{\cal M}_{mm} & = & (N_{\rm events} - 1)\bar{m}^2 \approx N_{\rm events} \bar{m}^2
\label{Eq35}
\eea
where only unique particle pairs are included in the summations. The variance normalization factor, $\bar{{\cal M}}_{mm}/\bar{{\cal S}}_{mm}$, for the diagonal bins is also equal to 2. The variance quantity is then given by
\bea
\langle (2{\cal S}_{mm} - {\cal M}_{mm})^2 \rangle & = &
\langle [ \sum_{j=1}^{N_{\rm events}} ({\cal S}_{mm}^j -\bar{m} -\mu_j) -
\sum_{j=1}^{N_{\rm events} - 1} {\cal M}_{mm}^{j,j+1} ]^2 \rangle
\nonumber \\
 &  & \hspace{-1.2in} = \frac{1}{4} \langle [2\sum_{j=1}^{N_{\rm events}} ({\cal S}_{mm}^j -\bar{m} -\mu_j)
- \sum_{j=1}^{N_{\rm events} - 1} ({\cal M}_{mm}^{j,j+1} + {\cal M}_{mm}^{j+1,j})]^2 \rangle
\label{Eq36}
\eea
where ${\cal M}_{mm}^{j,j+1} = {\cal M}_{mm}^{j+1,j}$. Separating the above into summations over even and odd event numbers, doubling the sibling-pair summation, assuming that $N_{\rm events} >> 1$ (in order that the contribution of missing mixed-events at the end of the event list is negligible), and using the noise reduction result from Ref.~\cite{TrainorReid} result in
\bea
\langle (2{\cal S}_{mm} - {\cal M}_{mm})^2 \rangle & = &
\langle [ \sum_{j=1}^{N_{\rm events}} (\mu_j^2 - \mu_j \mu_{j+1} - \bar{m} - \mu_j)]^2 \rangle
\nonumber \\
 & & \hspace{-0.5in} = \langle \sum_{j=1}^{N_{\rm events}} (\mu_j^4 + \mu_j^2 \mu_{j+1}^2 +\bar{m}^2 + \mu_j^2 - 2\bar{m}\mu_j^2 - 2\mu_j^3)
\nonumber \\
 & & \hspace{-0.5in}
+ \sum_{j \neq j^{\prime}} (\mu_j^2 \mu_{j^{\prime}}^2 - \bar{m}\mu_j^2 -\bar{m}\mu_{j^{\prime}}^2 + \bar{m}^2) \rangle .
\label{Eq37}
\eea
The second term vanishes in the $N_{\rm events} >> 1$ limit where
\bea
N_{\rm events}^2 (\sigma_{\mu}^4 - \bar{m} \sigma_{\mu}^2 -  \bar{m} \sigma_{\mu}^2 +\bar{m}^2)
 & \stackrel{\rm Poisson}{\longrightarrow} &    N_{\rm events}^2 (\bar{m}^2 - \bar{m}^2 - \bar{m}^2 + \bar{m}^2) \nonumber \\
 & = & 0. 
\label{Eq38}
\eea
The first term in Eq.~(\ref{Eq37}), using the moments of a Poisson distribution in Appendix B, simplifies to
\bea
\langle (2{\cal S}_{mm} - {\cal M}_{mm})^2 \rangle & = &
\nonumber \\
 & & \hspace{-1.0in}  N_{\rm events}
\langle \frac{1}{N_{\rm events}} \sum_{j=1}^{N_{\rm events}} (\mu_j^4 + \mu_j^2 \mu_{j+1}^2 +\bar{m}^2 + \mu_j^2 - 2\bar{m}\mu_j^2 - 2\mu_j^3) \rangle
\nonumber \\
 & & \hspace{-1.0in} \stackrel{\rm Poisson}{\longrightarrow}
N_{\rm events} \langle (3\bar{m}^2 + \bar{m}) + \bar{m}^2 + \bar{m}^2 + \bar{m} -2\bar{m}^2 - 2\bar{m} \rangle
= 3N_{\rm events} \bar{m}^2.
\nonumber \\
\label{Eq39}
\eea

The statistical error in diagonal bins for the single-event mixing algorithm is
\bea
\sqrt{(\Delta {\cal R}_{mm})^2} & = & \sqrt{\frac{3 N_{\rm events} \bar{m}^2}{(N_{\rm events} \bar{m}^2)^2}} = \frac{\sqrt{3}}{\bar{m}\sqrt{N_{\rm events}}}
\label{Eq40}
\eea
where the multiplicative factor in the numerator is $\sqrt{2}$ larger than that for the off-diagonal errors in Eq.~(\ref{Eq33}).

\subsection{Double-event mixing method}
\label{Sec:double}
In this method particles in each event $j$ are mixed with particles in events $j+1$ and $j+2$. The sibling and mixed-event pair summations, using the symbols ${\cal S}_{mn}^j$ and ${\cal M}_{mn}^{j,j^{\prime}}$ introduced above, and their limits are given by
\bea
{\cal S}_{mn} & = & \sum_{j=1}^{N_{\rm events}} {\cal S}_{mn}^j
\nonumber \\
\bar{{\cal S}}_{mn} & = & N_{\rm events} (\bar{m}\bar{n})
\label{Eq41} \\
{\cal M}_{mn} & = & \sum_{j=1}^{N_{\rm events} - 1} ({\cal M}_{mn}^{j,j+1} + {\cal M}_{mn}^{j+1,j})
+ \sum_{j=1}^{N_{\rm events} - 2} ({\cal M}_{mn}^{j,j+2} + {\cal M}_{mn}^{j+2,j})
\nonumber \\
\bar{\cal M}_{mn} & = &  2(N_{\rm events} - 1) (\bar{m}\bar{n}) + 2(N_{\rm events} - 2) (\bar{m}\bar{n})
\nonumber \\
 & \approx &  4N_{\rm events} (\bar{m}\bar{n}).
\label{Eq42}
\eea
The normalization factor for ${\cal R}_{mn}$ is approximately 4 and the variance to be calculated is
\bea
\langle (4{\cal S}_{mn} - {\cal M}_{mn})^2 \rangle & = &
\langle [ 4 \sum_{j=1}^{N_{\rm events}} {\cal S}_{mn}^j
- \sum_{j=1}^{N_{\rm events} - 1} ({\cal M}_{mn}^{j,j+1} + {\cal M}_{mn}^{j+1,j})
\nonumber \\
 & & - \sum_{j=1}^{N_{\rm events} - 2} ({\cal M}_{mn}^{j,j+2} + {\cal M}_{mn}^{j+2,j})]^2 \rangle .
\label{Eq43}
\eea

By writing out the sum over sibling pairs four times it is found that the resulting summations in Eq.~(\ref{Eq43}) can be separated into sums over pairs of events where arbitrary event $j$ is mixed with event $j+k$, and in this example $k=1$ and 2. For further brevity the symbol $\Delta_{mn}^{j,k}$ is introduced which is defined by
\bea
\Delta_{mn}^{j,k} & \equiv & {\cal S}_{mn}^j + {\cal S}_{mn}^{j+k} - {\cal M}_{mn}^{j,j+k} - {\cal M}_{mn}^{j+k,j}.
\label{Eq43b}
\eea
After rearranging the sums the above variance in the $N_{\rm events} >> 1$ limit can be expressed as
\bea
\langle (4{\cal S}_{mn} - {\cal M}_{mn})^2 \rangle & = &
\langle [ \sum_{j=2,{\rm even}} \Delta_{mn}^{j,1} 
        + \sum_{j=1,{\rm odd}}  \Delta_{mn}^{j,1}
\nonumber \\
 & & \hspace{-0.5in}
        + \sum_{j=1,4,7,\cdots}  \Delta_{mn}^{j,2}
        + \sum_{j=2,5,8,\cdots}  \Delta_{mn}^{j,2}
        + \sum_{j=3,6,9,\cdots}  \Delta_{mn}^{j,2}  ]^2 \rangle
\label{Eq44}
\eea
where, as in Ref.~\cite{TrainorReid},
\bea
\Delta_{mn}^{j,k} & = & 
\mu_j \nu_j + \mu_{j+k} \nu_{j+k} - \mu_j \nu_{j+k} - \mu_{j+k} \nu_j .
\label{Eq45}
\eea
Substituting $\Delta_{mn}^{j,k}$ into Eq.~(\ref{Eq44}) and recombining the summations gives
\bea
\langle (4{\cal S}_{mn} - {\cal M}_{mn})^2 \rangle & = &
\langle [ \sum_{j=1}^{N_{\rm events}} \left( 4\mu_j \nu_j
- (\mu_j \nu_{j+1} +  \mu_{j+1} \nu_j) \right.
\nonumber \\
 & &
 \left.  - (\mu_j \nu_{j+2} +  \mu_{j+2} \nu_j) \right) ]^2 \rangle.
\label{Eq46}
\eea
Eliminating cross terms involving products of fluctuations between different events results in 
\bea
\langle (4{\cal S}_{mn} - {\cal M}_{mn})^2 \rangle & = &
\langle \sum_{j=1}^{N_{\rm events}} (16 \mu_j^2 \nu_j^2 + \mu_j^2 \nu_{j+1}^2
+ \mu_{j+1}^2 \nu_j^2 + \mu_j^2 \nu_{j+2}^2 + \mu_{j+2}^2 \nu_j^2) \rangle
\nonumber \\
 & = & 20 N_{\rm events} \langle \sigma_{\mu}^2 \sigma_{\nu}^2 \rangle
 \stackrel{\rm Poisson}{\longrightarrow} 20 N_{\rm events} (\bar{m}\bar{n}).
\label{Eq47}
\eea

The statistical error in off-diagonal bins for the double-event mixing algorithm is given by
\bea
\sqrt{(\Delta{\cal R}_{mn})^2} & = & \sqrt{\frac{20 N_{\rm events} \bar{m}\bar{n}}{(4N_{\rm events} \bar{m}\bar{n})^2}} = \frac{\sqrt{5/4}}{\sqrt{N_{\rm events} \bar{m}\bar{n}}}
\label{Eq48}
\eea
which is smaller than the above, single-event mixing error and is $\sqrt{5/8}$ smaller than the error resulting from the Reid-Trainor method. Again, this error equals $\sqrt{\bar{\cal S}_{mn}^{-1} + \bar{\cal M}_{mn}^{-1}}$. The error for diagonal bins is obtained as in the above sections and is given by
\bea
\sqrt{(\Delta{\cal R})_{mm}^2} & = & \frac{\sqrt{5/2}}{\bar{m}\sqrt{N_{\rm events}}} .
\label{Eq49}
\eea

\subsection{Multiple-event mixing method}
\label{Sec:multi}

The preceding event-mixing methods can be readily extended to an arbitrary number of mixing events $K = 1,2,3,\cdots$. In the following derivation however we continue to assume that $N_{\rm events} >> K$. The sums of sibling and mixed-event particle pairs in off-diagonal bins and their limits are
\bea
{\cal S}_{mn} & = & \sum_{j=1}^{N_{\rm events}} {\cal S}_{mn}^j
\nonumber \\
\bar{\cal S}_{mn} & = & N_{\rm events}(\bar{m}\bar{n})
\label{Eq50} \\
{\cal M}_{mn} & = & \sum_{j=1}^{N_{\rm events} -1}({\cal M}_{mn}^{j,j+1} + {\cal M}_{mn}^{j+1,j})
+ \sum_{j=1}^{N_{\rm events} -2}({\cal M}_{mn}^{j,j+2} + {\cal M}_{mn}^{j+2,j})
\nonumber \\
 & + & \cdots + \sum_{j=1}^{N_{\rm events} -K}({\cal M}_{mn}^{j,j+K} + {\cal M}_{mn}^{j+K,j})
\nonumber \\
\bar{\cal M}_{mn} & \approx &  2KN_{\rm events}(\bar{m}\bar{n})
\label{Eq51}
\eea
and using the definition in Eq.~(\ref{Eq43b}) the variance is given by
\bea
\langle (2K{\cal S}_{mn} - {\cal M}_{mn})^2 \rangle & = &
\langle [ \sum_{k=1}^{K} \sum_{k^{\prime}=1}^{k+1} 
\sum_{\ell \geq 0}
\Delta_{mn}^{k^{\prime}+\ell (k+1),k} ]^2 \rangle
\label{Eq52}
\eea
where event index $j = k^{\prime}+\ell (k+1)$, $\ell = 0,1,2,\cdots$. Recombining the summations over sibling pair quantities $(\mu_j \nu_j + \mu_{j+k} \nu_{j+k})$ results in $2K\sum_{j=1}^{N_{\rm events}} \mu_j \nu_j$. The summations over the last two terms, $(\mu_j \nu_{j+k} + \mu_{j+k} \nu_{j})$, include each mixed-event only once and may be simplified to $\sum_{k=1}^{K} \sum_j (\mu_j \nu_{j+k} + \mu_{j+k} \nu_{j})$. The variance can then be expressed as
\bea
\langle (2K{\cal S}_{mn} - {\cal M}_{mn})^2 \rangle & = &
\langle [ \sum_{j=1}^{N_{\rm events}} (2K\mu_j \nu_j - \sum_{k=1}^{K}
(\mu_j \nu_{j+k} + \mu_{j+k} \nu_{j}))]^2 \rangle .
\label{Eq53}
\eea
As before the average over cross terms between different events vanishes resulting in
\bea
\langle (2K{\cal S}_{mn} - {\cal M}_{mn})^2 \rangle & = &
\langle \sum_{j=1}^{N_{\rm events}} (4K^2 \mu_j^2 \nu_j^2 + \sum_{k=1}^{K} (
\mu_j^2 \nu_{j+k}^2 + \mu_{j+k}^2 \nu_{j}^2)) \rangle
\nonumber \\
 &  & \hspace{-1.0in} =  N_{\rm events} \langle(4K^2+2K)\sigma_{\mu}^2 \sigma_{\nu}^2 \rangle
\stackrel{\rm Poisson}{\longrightarrow} (4K^2+2K) N_{\rm events}(\bar{m}\bar{n})
\label{Eq54}
\eea
and the error in off-diagonal bins is
\bea
\sqrt{(\Delta {\cal R}_{mn})^2} & = & \sqrt{\frac{(4K^2+2K)N_{\rm events}\bar{m}\bar{n}}
{(2KN_{\rm events}\bar{m}\bar{n})^2}} = \sqrt{ \frac{1+\frac{1}{2K}}{N_{\rm events}\bar{m}\bar{n}}}.
\label{Eq55}
\eea 
Evaluating this result for $K = 1$ and 2 gives the above errors for single-event and double-event mixing in Eqs.~(\ref{Eq33}) and (\ref{Eq48}), respectively. The above error equals $\sqrt{\bar{\cal S}_{mn}^{-1} + \bar{\cal M}_{mn}^{-1}}$.

To summarize, for each event-mixing algorithm presented here the sibling minus mixed-event pair summations can be accurately rearranged into sums over event-pair-wise differences between sibling and mixed-event pairs as in Ref.~\cite{TrainorReid}, resulting in significant statistical noise reduction. The statistical errors for the correlation quantity $({\cal R}_{mn} - 1)$ are determined by the total numbers of sibling and mixed-event pairs of particles in 2D bin $(m,n)$, given by $\sqrt{\bar{\cal S}_{mn}^{-1} + \bar{\cal M}_{mn}^{-1}}$.

\subsection{Prefactor, charge dependence and pair projections}
\label{Sec:prefactor}

The correlation quantity $({\cal R}_{mn} - 1)$ in Eq.~(\ref{Eq5}) represents the event-averaged number of correlated particle pairs per final-state particle pair produced in the collisions. In some analyses~\cite{AxialCI,LHC1,LHC2} the authors reported the number of correlated particle pairs per final-state particle in which case $({\cal R}_{mn} - 1)$ was scaled by a prefactor~\cite{AxialCI,LHC1,LHC2} proportional to total event-multiplicity. For example, in the analysis of Au + Au collisions at $\sqrt{s_{NN}}$ = 200~GeV~\cite{AxialCI} the prefactor was computed using efficiency corrected, charged-particle yields within the detector acceptance defined by
\bea
PF_{\rm CI}^{(\eta,\phi)} & = & \left[ \frac{d^2N_{\rm chrg}}{d\eta_1 d\phi_1}
\frac{d^2N_{\rm chrg}}{d\eta_2 d\phi_2} \right]^{1/2}  = \frac{d^2N_{\rm chrg}}{d\eta d\phi}
\label{Eq56}
\eea
for correlations on 2D angular space $(\eta_1 - \eta_2,\phi_1 - \phi_2)$ for arbitrary particle pair (1,2) and at mid-rapidity where $d^2N_{\rm chrg}/d\eta d\phi$ is approximately constant. Subscript ``CI'' (charge independent) means that all charge-sign pairs are included, i.e. $++$, $--$, $+-$, and $-+$.  Other kinematic projections are possible. In Ref.~\cite{LizYtYt} correlations on 2D transverse momentum space were presented with corresponding prefactor
\bea
PF_{\rm CI}^{(p_t,p_t)} & = & \left[ \frac{d^2N_{\rm chrg}}{dp_{t1} d\eta_1}
\frac{d^2N_{\rm chrg}}{dp_{t2} d\eta_2}  \right]^{1/2}.
\label{Eq57}
\eea
Statistical errors in the prefactor could be included in the final, reported correlation quantities but this factor is usually treated as a simple scaling of the correlation measurement.

Insight into the dynamics underlying the observed correlations can be gained by making further selections on the particle pairs. For example, correlations for positive versus negative charged particle pairs, or like-charge-sign pairs ($++$ and $--$) versus unlike-sign pairs ($+-$ and $-+$), can help differentiate charge-ordering effects~\cite{chargeorder1,chargeorder2} in hadronization, resonance decay contributions, quantum correlations for identical particles~\cite{HBT}, and photon pair-production contamination. Selection of low-$p_t$ or high-$p_t$ particles can help differentiate soft and hard-scattering processes~\cite{AxialCI}. Similarly, for correlations on 2D transverse momentum-space particle pairs with smaller or larger relative azimuthal angles, e.g. $|\phi_1 - \phi_2| \leq \pi/2$ (near-side) or $|\phi_1 - \phi_2| >  \pi/2$ (away-side), in conjunction with charge-pair selections, can help differentiate the $p_t$ structures of near-side, jet-like fragmentation from that of away-side, or back-to-back dijet fragmentation.  

If the prefactor, which is the square-root of a pair density, is computed using the same charge-sign and/or kinematic selections used in constructing the correlation $({\cal R}_{mn} - 1)$, then the statistical error in the final quantity, $PF({\cal R}_{mn} - 1)$, is unaffected by those selections. This is a result of the noise reduction inherent in the event-mixing methods presented here and in Ref.~\cite{TrainorReid} where the statistical error is inversely proportional to the square-root of the number of particle pairs in the bin, $(N_{\rm events}\bar{m}\bar{n})$. For example, if only like-sign pairs are selected the total number of pairs is reduced by 1/2 if the numbers of positive and negative particles are the same (approximately true for relativistic heavy-ion collisions). The prefactor is therefore reduced by $1/\sqrt{2}$ while the error in each bin is increased by $\sqrt{2}$. Similarly, selecting only near-side or away-side particle pairs reduces the number of pairs by 1/2 which again decreases the prefactor by $1/\sqrt{2}$ and increases the bin errors by $\sqrt{2}$. In general, such pair selections do not affect the prefactor scaled statistical errors studied here. This result is summarized as follows:
\bea
\sqrt{(\Delta\{PF({\cal R}_{mn} - 1)\})_{\rm arb.-pair-selection}^2}
 & = &
PF_{\rm CI} \sqrt{(\Delta{\cal R}_{mn})^2}
\label{Eq58}
\eea
where the left-hand-side represents the root-mean-square statistical error of quantity $PF({\cal R}_{mn} - 1)$ for arbitrary charge-sign and/or azimuth angle pair selections. Eq.~(\ref{Eq58}) applies to each event-mixing method discussed here.

\section{Monte Carlo simulations}
\label{Sec:montecarlo}

Throughout the preceding derivations it was assumed in the final steps that the frequency distributions for event-wise particle number in bins $m$ and $n$ were consistent with Poisson distributions, i.e. variance $\sigma_{\mu}^2$ equaled mean $\bar{m}$. In practical data analysis, collision events are usually grouped into subsets based on global event properties such as total multiplicity in the detector acceptance~\cite{AxialCI,MikeThesis,LizThesis}. Clearly the multiplicity range chosen for the event collections can affect the frequency distributions in the $m$ and $n$ bins.  In addition, if the $m$,$n$ bin sizes are comparable to the acceptance, then the multiplicity distribution of the event collection will also affect that in the bins. In this section Monte Carlo simulations are used to determine the differences between actual event-wise particle number frequency distributions in the bins and a Poisson distribution. If the distributions differ significantly from a Poisson, then the actual variances should be used in the above derivations.

In Au + Au collisions at energies of 200~GeV per colliding nucleon + nucleon pair over a thousand charged particles can be produced at mid-rapidity within the tracking acceptance of the STAR detector~\cite{STAR} ($p_t > 0.15$~GeV/$c$, $|\eta| \leq 1$ and $2\pi$ in azimuth). For events selected with a minimum-bias trigger~\cite{AxialCI} the multiplicity frequency distribution for total, observed multiplicity, $N_{\rm chrg}$, is approximately proportional to $N_{\rm chrg}^{-3/4}$ except near the lower and upper multiplicity end-points~\cite{TomPowerLaw,MCGpaper}. The events are then subdivided into multiplicity classes or centrality. For example, the multiplicity ranges for peripheral, intermediate (mid-central), and head-on (most-central) collisions from Ref.~\cite{MikeThesis} are [15,35], [152,187] and [952,1002], respectively.

For the present example, correlations on 2D transverse momentum space were considered where the kinematic variable used is transverse rapidity, defined by $y_t = \log[(m_t + p_t)/m_0]$, $m_t^2 = p_t^2 + m_0^2$ and for non-identified charged particles $m_0$, which regulates the singularity in $y_t$ at $p_t = 0$, was assumed to be the pion mass, 0.14~GeV/$c^2$. The transverse rapidity spectrum for each multiplicity bin was obtained from the efficiency and contamination corrected $p_t$ spectrum data~\cite{STARspectra,PHENIXspectra} and is accurately represented by a Levy distribution given by
\bea
\frac{d^2N_{\rm chrg}}{dy_td\eta} & = & \frac{2\pi p_t m_t A}
{[1 + \beta(m_t - m_0)/n]^n}.
\label{Eq59}
\eea
Parameters $T = 1/\beta$ and $n$ from data fitting for peripheral, mid-central and most-central collisions are (0.154~GeV, 10.42), (0.199~GeV, 13.32), and (0.226~GeV, 17.55), respectively. Amplitude $A$ is determined by event-wise multiplicity. In the correlation analysis particle pairs were binned on 2D $(y_{t1},y_{t2})$ space where $y_t \in [1.0,4.5]$ with $25 \times 25$ 2D $(m,n)$ bins of equal size.

In the simulations, the power-law distribution $N_{\rm chrg}^{-3/4}$ was randomly sampled within each of the above multiplicity ranges, the single-particle distribution in Eq.~(\ref{Eq59}) was sampled for the event-wise multiplicity, the 25 single-particle $y_t$ bins were filled, and the number of particles in each $y_t$ bin for each event was recorded. The particle number frequency distributions for the 25 $y_t$ bins for the simulated events were compared with Poisson distributions. This was done by calculating the ratio of the computed variance, $\sigma_{\rm MC}^2$, to the variance for a Poisson distribution, $\sigma_{\rm Poisson}^2 = \bar{n}(y_t)$, where $\bar{n}(y_t)$ is the mean particle number in the $y_t$ bin. The resulting ratios were compared with unity. The uncertainty in the ratio of variances for each $y_t$ bin was estimated by repeating the above simulation procedure many times, corresponding to statistically independent collections of events. The mean and width of the resulting distribution of variance ratios were computed. 

\begin{figure}
\includegraphics[keepaspectratio,width=4.0in]{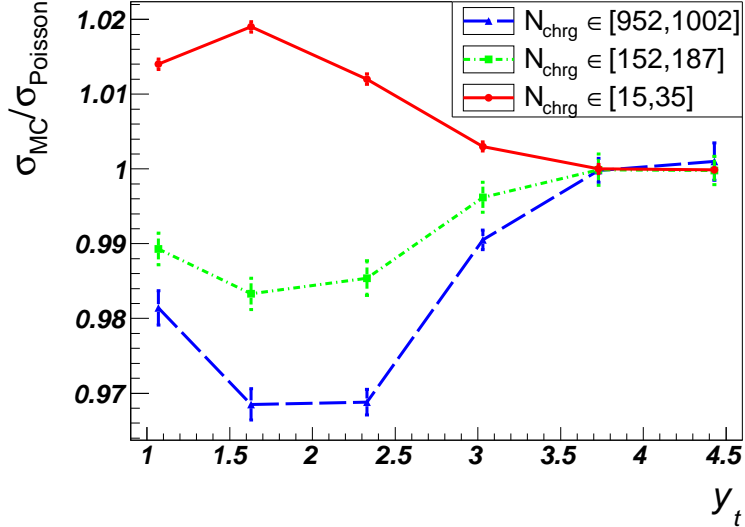}
\caption{\label{Figure1}
(Color online) Ratios $\sigma_{\rm MC}/\sigma_{\rm Poisson}$ as described in the text and listed in Table~\ref{Table1} for three event-multiplicity ranges where the solid, dashed-dotted and dashed lines connecting the points correspond to peripheral, mid-central and most-central collisions, respectively.}
\end{figure}

Results for $\sigma_{\rm MC}/\sigma_{\rm Poisson}$ for six of the twenty-five $y_t$ bins are listed in Table~\ref{Table1} and shown in Fig.~\ref{Figure1} for the three event-multiplicity ranges listed above. For the peripheral, mid-central and most-central collisions $10^6$, $10^5$ and $10^5$ events were generated, respectively. One-hundred event collections were simulated for each centrality.  The calculations provided sufficient statistical accuracy to determine a significant difference between $\sigma_{\rm MC}/\sigma_{\rm Poisson}$ and unity for most $y_t$ bins. In all cases $\sigma_{\rm MC}$ agreed with the Poisson limit $\sqrt{\bar{n}(y_t)}$ to within 3\% or less, thus confirming the Poisson assumption for the present application to Au + Au collisions at 200~GeV.

\begin{centering}
\begin{table*}[t]
\caption{Ratios of standard deviations of simulated particle-number frequency distributions in selected $y_t$ bins ($\sigma_{\rm MC}$) to standard deviations assuming Poisson distributions as explained in the text. The first column lists the range in transverse rapidity for the selected $y_t$ bins and the remaining three columns list the ratios and their statistical errors for three event-multiplicity ranges.}
\label{Table1}
\begin{tabular}{|c|ccc|}
\hline \hline
      &  \multicolumn{3}{c|}{$\sigma_{\rm MC}/\sigma_{\rm Poisson}$}  \\
\hline
$y_t$ range & $N_{\rm chrg} \in [15,35]$ & [152,187] & [952,1002] \\
\hline
1.00-1.14  &  1.014$\pm$0.0007 & 0.9893$\pm$0.0021 & 0.9814$\pm$0.0023 \\
1.56-1.70  &  1.019$\pm$0.0007 & 0.9833$\pm$0.0021 & 0.9685$\pm$0.0021 \\
2.26-2.40  &  1.012$\pm$0.0007 & 0.9854$\pm$0.0023 & 0.9688$\pm$0.0017 \\
2.96-3.10  &  1.003$\pm$0.0006 & 0.9962$\pm$0.0020 & 0.9905$\pm$0.0013 \\
3.66-3.80  &  1.000$\pm$0.0006 & 0.9999$\pm$0.0021 & 0.9998$\pm$0.0016 \\
4.36-4.50  &  0.9999$\pm$0.0003 & 0.9998$\pm$0.0019 & 1.001$\pm$0.0025 \\
\hline \hline
\end{tabular}
\end{table*}
\end{centering}

Monte Carlo simulations were also used to verify the noise reduction predicted in Ref.~\cite{TrainorReid} and for the derivations in Sections~\ref{Sec:general} and \ref{Sec:other}. The simulations were also used to check the error scaling factors $\sqrt{2}$, $\sqrt{3/2}$ and $\sqrt{5/4}$ for the Reid-Trainor, single- and double-event mixing algorithms. In Ref.~\cite{TrainorReid} and in Appendix C the statistical error for correlations obtained with mixed-event particle pairs taken from a different set of events than those used for the sibling particle pairs is shown to be proportional to $[(\bar{m}^{-1} + \bar{n}^{-1} + (\bar{m}\bar{n})^{-1})/N_{\rm events}]^{1/2}$. This error depends on particle number and therefore differs significantly from the errors derived in Ref.\cite{TrainorReid} and in Sections~\ref{Sec:general} and \ref{Sec:other} which are proportional to $(N_{\rm events}\bar{m}\bar{n})^{-1/2}$. Finally, the simulations were used to verify that the errors in 2D diagonal bins are larger than those in off-diagonal bins by a factor of $\sqrt{2}$.

As in the preceding simulation the $N_{\rm chrg}^{-3/4}$ frequency distribution was randomly sampled to obtain an event multiplicity, the $y_t$ distribution in Eq.~(\ref{Eq59}) was sampled, and the 25 single-particle $y_t$ bins were filled. Sibling and mixed-event pairs were counted for each 2D $(m,n)$ bin on $(y_{t1},y_{t2})$ space using the event-mixing algorithms in Ref.~\cite{TrainorReid} (see Sec.~\ref{Sec:general}), the single-event mixing described in Sec.~\ref{Sec:single}, or the double-event mixing described in Sec.~\ref{Sec:double}. The correlation quantity $({\cal R}_{mn} - 1)$ was computed for all $(m,n)$ bins using all simulated events in the collection. This entire process was repeated many times thus generating independent, statistically equivalent event collections and corresponding values for $({\cal R}_{mn} - 1)$. From these distributions of $({\cal R}_{mn} - 1)$ the means and widths in each $(m,n)$ bin were obtained, including those for on-diagonal and off-diagonal bins. The resulting Monte Carlo uncertainties are directly comparable to the analytic results for statistical errors in Eqs.~(\ref{Eq10}), (\ref{Eq17}), (\ref{Eq33}), (\ref{Eq40}), (\ref{Eq48}), and (\ref{Eq49}). Results are reported below for the mid-central multiplicity bin with $N_{\rm chrg} \in [152,187]$ and for two diagonal $(m,m)$ bins on $(y_{t1},y_{t2})$ defined by ranges ([1.56,1.70],[1.56,1.70]) and ([2.96,3.10],[2.96,3.10]), and for the corresponding off-diagonal $(m,n)$ bin defined by ([1.56,1.70],[2.96,3.10]). The mean, single-particle numbers in these two $y_t$ bins are 14.3 and 3.39, respectively.

The scaling of the error with particle pair number, rather than with single particle number can be checked by comparing the simulated errors for two diagonal bins. For the above diagonal bins the ratios of the analytical errors in the lower to upper $y_t$ bins are 0.237 for particle-pair scaling (i.e. with noise reduction) and 0.462 for single-particle scaling (statistically independent events for sibling and mixed-event particle-pairs, see Appendix C). The simulated ratios for the Reid-Trainor, single-event mixing, and double event-mixing algorithms are 0.241, 0.226 and 0.236, respectively, which agree with the noise reduced, particle-pair scaling. 

The scaling factors for the errors can be checked by comparing the analytic ratio for single-event mixing to Reid-Trainor mixing, equal to $\sqrt{3/2}/\sqrt{2} = 0.866$, and the ratio for double-event mixing to Reid-Trainor mixing, equal to $\sqrt{5/4}/\sqrt{2} = 0.791$, to the corresponding ratios from the simulations. The Monte Carlo error results for the lower and upper $y_t$ diagonal bins are 0.851 and 0.911 (ratio of single-event mixing to Reid-Trainor mixing) and 0.785 and 0.803 (ratio of double-event mixing to Reid-Trainor mixing), respectively. Both pairs of Monte Carlo results are in good agreement with the corresponding analytic expressions.

Finally, the $\sqrt{2}$ increase in the diagonal bin errors can be checked via the ratio $E(m,n)^2/(E(m,m)E(n,n))$ where $E(m,n)$ is the error in bin $(m,n)$. The analytical value of this ratio equals 1/2 for the three event-mixing algorithms considered in this section. The simulation results for the Reid-Trainor, single-event mixing, and double-event mixing algorithms using the above bins are 0.52, 0.52 and 0.50, respectively, in good agreement with the analytic errors.

The simulated and analytic absolute errors agree within 2\%-6\%, where the few percent effects of non-Poisson distributions discussed above contribute here as well. The scaling factors discussed above and the differences between the errors obtained with particle-pair number scaling versus single-particle number scaling are much larger than these few percent differences. Overall, the simulations confirm the analytic expressions. The Poisson distribution assumption should be checked for each application.

\section{Conclusions}
\label{Sec:conclusions}
The event-mixing and error propagation method of Reid-Trainor~\cite{TrainorReid} for a multiplicity-ordered event collection, grouped into pairs of events for particle-pair mixing, was generalized to include event-mixing groups with arbitrary number of events. The method was then applied to other event-mixing algorithms which have been used in correlation analysis of relativistic heavy-ion collision data. The noise reduction arising from the partial cancellation of event-wise single-particle multiplicity fluctuations was shown to occur for those other methods in which particle-pair mixing is done between each event in the collection and the next 1, 2 or $K$ (where $K << N_{\rm events}$) events in the list. 

The statistical errors for analyses which use the event-mixing group method of Ref.~\cite{TrainorReid} are given in Eqs.~(\ref{Eq10}) and (\ref{Eq17}) for two events per mixing group and in Eq.~(\ref{Eq25}) for $\epsilon > 1$ events in each event-mixing group. Errors for correlation analysis based on single-event mixing and double-event mixing are given in Eqs.~(\ref{Eq33}), (\ref{Eq40}) and Eqs.~(\ref{Eq48}), (\ref{Eq49}), respectively. Correlation errors for all event-mixing algorithms in which each event is mixed with the next $K$ events in the list are given in Eq.~(\ref{Eq55}). Statistical errors for correlations reported as the number of correlated pairs per final-state particle, including charge and/or kinematic dependent pair selections~\cite{AxialCI,LHC1,LHC2,MikeThesis,LizThesis}, are given in Eq.~(\ref{Eq58}).

The statistical errors in correlations for each of the event-mixing algorithms studied here using the Reid-Trainor method are inversely proportional to the square-root of the number of pairs of particles in a bin. This result is in contrast to that obtained when the two-particle reference distribution is constructed from a set of events independent from those used for sibling pairs. In the latter case the errors are inversely proportional to the square-root of the number of particles in the bin and can be much larger than the errors which follow from the event-mixing methods discussed here and in Ref.~\cite{TrainorReid}. As long as the sum over sibling particle pairs minus mixed-event particle pairs can be accurately approximated by sibling minus mixed-event particle pair differences summed over pairs of events as in Ref.~\cite{TrainorReid} and in this paper, then the Reid-Trainor error reduction will follow and the correlation errors will be inversely proportional to the square-root of the total, bin-wise number of particle pairs. It should also be noted that error propagation algorithms based on bin-wise single-particle counts which neglect covariance $\Delta({\cal S},{\cal M})_{mn}$ in Eq.~(\ref{Eq6}) will produce erroneous results.

Monte Carlo simulations were employed to test the accuracy of the Poisson distribution approximations used in the error calculations. Simulations were also used to verify the predicted reduction in statistical errors for the Reid-Trainor, single-event and double-event mixing algorithms. Finally, the analytic error results presented here may be helpful in optimizing correlation data analysis when more aggressive event-mixing algorithms are considered and the benefit from reduced statistical errors needs to be weighed against the additional computational cost.

\vspace{0.2in}
{\bf Acknowledgements}
\vspace{0.1in}

The authors would like to thank Professor Tom Trainor of the Univ. of Washington for many informative discussions relevant to this work. This research was supported in part by the Office of Science of the U. S. Department of Energy under grants DE-FG02-94ER40845 and DE-SC0013391.

\section{Appendix A}

In deriving Eq.~(\ref{Eq8}) event averages and cross terms involving averages over pairs of different events must be calculated. The necessary steps are given here. Similar calculations are required for the other event-mixing algorithms described in this paper. Starting with the first line in Eq.~(\ref{Eq8}), replacing the group summation in the first part of Eq.~(\ref{Eq8}) with event summations, expanding the quadratic term, and retaining the non-vanishing averages yield the following:
\bea
\langle ({\cal S}_{mn} - {\cal M}_{mn})^2 \rangle & = &
\langle [ \sum_{j=1}^{N_{\rm events}} \mu_j \nu_j  -
\sum_{j=1,{\rm odd}}^{N_{\rm events}} (\mu_j \nu_{j+1} + \mu_{j+1} \nu_j)]^2 \rangle
\nonumber \\
 & = & \langle [ \sum_{j=1,{\rm odd}}^{N_{\rm events}} (\mu_j \nu_j - \mu_j \nu_{j+1} )
+ \sum_{j=2,{\rm even}}^{N_{\rm events}} (\mu_j \nu_j - \mu_j \nu_{j-1} ) ]^2 \rangle
\nonumber \\
 & = & \langle \sum_{j=1,{\rm odd}}^{N_{\rm events}} (\mu_j \nu_j - \mu_j \nu_{j+1} )^2
+  \sum_{j=2,{\rm even}}^{N_{\rm events}}  (\mu_j \nu_j - \mu_j \nu_{j-1} )^2
\nonumber \\
 & + & \sum_{j \neq j^{\prime},{\rm odd}} (\mu_j \nu_j - \mu_j \nu_{j+1} )
(\mu_{j^{\prime}} \nu_{j^{\prime}} - \mu_{j^{\prime}} \nu_{{j^{\prime}}+1} )
\nonumber \\
 & + & \sum_{j \neq j^{\prime},{\rm even}} (\mu_j \nu_j - \mu_j \nu_{j-1} )
(\mu_{j^{\prime}} \nu_{j^{\prime}} - \mu_{j^{\prime}} \nu_{{j^{\prime}}-1} )
\nonumber \\
 & + & 2 \sum_{j=1,{\rm odd}}^{N_{\rm events}}
\sum_{j^{\prime}=2,{\rm even}}^{N_{\rm events}}
(\mu_j \nu_j - \mu_j \nu_{j+1} ) 
(\mu_{j^{\prime}} \nu_{j^{\prime}} - \mu_{j^{\prime}} \nu_{{j^{\prime}}-1} )
\rangle
\nonumber \\
 & = & \langle \sum_{j=1}^{N_{\rm events}} \mu_j^2 \nu_j^2 
+ \sum_{j=1,{\rm odd}}^{N_{\rm events}} \mu_j^2 \nu_{j+1}^2
+ \sum_{j=2,{\rm even}}^{N_{\rm events}} \mu_j^2 \nu_{j-1}^2 \rangle .
\nonumber 
\eea
The last line is equivalent to Eq.~(\ref{Eq8}). Event collection averages over cross terms involving pairs of different events, for example
\bea
 & &  \langle \sum_{j=1,{\rm odd}}^{N_{\rm events}} \mu_j^2 \nu_j \nu_{j+1} \rangle , \hspace{0.1in}
 \langle  \sum_{j \neq j^{\prime},{\rm odd}} \mu_j \nu_j \mu_{j^{\prime}} \nu_{j^{\prime}} \rangle , \hspace{0.1in}
\langle  \sum_{j \neq j^{\prime},{\rm odd}} \mu_j \nu_j \mu_{j^{\prime}} \nu_{j^{\prime}+1} \rangle \hspace{0.1in} {\rm and}
\nonumber \\
 & &  \langle  \sum_{j=1,{\rm odd}}^{N_{\rm events}}
\sum_{j^{\prime}=2,{\rm even}}^{N_{\rm events}}
\mu_j \nu_j \mu_{j^{\prime}} \nu_{j^{\prime}} \rangle ,
\nonumber
\eea
vanish because bin-wise multiplicity fluctuations are uncorrelated from event-to-event.  Cross term
\bea
\langle  \sum_{j=1,{\rm odd}}^{N_{\rm events}}
\sum_{j^{\prime}=2,{\rm even}}^{N_{\rm events}}
\mu_j \nu_j \mu_{j^{\prime}} \nu_{j^{\prime}-1} \rangle & &
\nonumber \\
 &  & \hspace{-1.0in} = \langle
\sum_{j=1,{\rm odd}}^{N_{\rm events}}
\sum_{\stackrel{j^{\prime}=2,{\rm even}}{j^{\prime} \neq j+1}}^{N_{\rm events}}
\mu_j \nu_j \mu_{j^{\prime}} \nu_{j^{\prime}-1} +
\sum_{j=1,{\rm odd}}^{N_{\rm events}}
\mu_j \mu_{j+1} \nu_j^2 \rangle
\nonumber
\eea
also vanishes for the same reason where the $j^{\prime} = j+1$ term is separated out.

\section{Appendix B}

Moments of the multiplicity frequency distribution in arbitrary bin $m$ involve averages of the event-wise fluctuations
\bea
\frac{1}{N_{\rm events}} \sum_{j=1}^{N_{\rm events}} \mu_j^p
\nonumber
\eea
where exponent $p = 0,1,2,\cdots$ and $\mu_j = m_j - \bar{m}$, where $m_j$ is the multiplicity in bin $m$ for event $j$.  This average can be expressed as
\bea
\frac{1}{N_{\rm events}} \sum_{j=1}^{N_{\rm events}} (m_j - \bar{m})^p & = &
\sum_M \frac{N(M)}{N_{\rm events}} \frac{1}{N(M)} \sum_{j=1}^{N(M)}
[(m_j - \bar{m})^p]|_{m_j = M}  \nonumber \\
 & = & \sum_M \frac{N(M)}{N_{\rm events}} (M-\bar{m})^p
\nonumber
\eea
where $N(M)$ is the number of events with $M$ particles in bin $m$ and $\sum_{j=1}^{N(M)}$ sums over all events with $m_j = M$. For Poisson distributions $N(M)/N_{\rm events} = {\cal P}(M,\bar{m})$ where ${\cal P}(M,\bar{m}) = \bar{m}^M e^{-\bar{m}}/M!$. The moments are calculated via the summations given by
\bea
\frac{1}{N_{\rm events}} \sum_{j=1}^{N_{\rm events}} (m_j - \bar{m})^p & = &
\sum_M {\cal P}(M,\bar{m}) (M-\bar{m})^p.
\nonumber
\eea
After expanding the above polynomial the resulting Poisson weighted sums can be related to Bell polynomials~\cite{Bellpoly}, defined by
\bea
B_p(\bar{m}) & = & \sum_{M=0}^{\infty} M^p {\cal P}(M,\bar{m})
\nonumber
\eea
where
\bea
B_0(\bar{m}) & = & 1 \nonumber \\
B_1(\bar{m}) & = & \bar{m} \nonumber \\
B_2(\bar{m}) & = & \bar{m}^2 + \bar{m} \nonumber \\
B_3(\bar{m}) & = & \bar{m}^3 + 3\bar{m}^2 + \bar{m} \nonumber \\
B_4(\bar{m}) & = & \bar{m}^4 + 6\bar{m}^3 + 7\bar{m}^2 + \bar{m} \nonumber 
\eea
and so on. The first few values of $\sum_M {\cal P}(M,\bar{m}) (M-\bar{m})^p$ for $p \in [0,4]$ are 1, 0, $\bar{m}$, $\bar{m}$, $(3\bar{m}^2+\bar{m})$, respectively.

\section{Appendix C}

The statistical error for correlation quantity $({\cal R}_{mn} - 1)$ when independent sets of events are used to calculate the sibling and mixed-event pair sums is derived here for the event-mixing algorithm in Ref.~\cite{TrainorReid}. As before
\bea
{\cal S}_{mn} & = & \sum_{j=1}^{N_{\rm events}} (\bar{m}+\mu_j)(\bar{n}+\nu_j)
\nonumber \\
\bar{\cal S}_{mn} & = &
N_{\rm events}(\bar{m}\bar{n}) \nonumber \\
{\cal M}_{mn} & = & \sum_{j^{\prime}=1,{\rm odd}}^{N_{\rm events}}
[(\bar{m}+\mu_{j^{\prime}}) (\bar{n}+\nu_{j^{\prime}+1}) + 
 (\bar{m}+\mu_{j^{\prime}+1}) (\bar{n}+\nu_{j^{\prime}}) ]
\nonumber \\
\bar{\cal M}_{mn} & = & N_{\rm events}(\bar{m}\bar{n}) \nonumber
\eea
where different event collections $\{j\}$ and $\{j^{\prime}\}$ are statistically equivalent. The error of $({\cal R}_{mn} - 1)$ squared equals $\langle ({\cal S}_{mn} - {\cal M}_{mn})^2 \rangle / \bar{{\cal M}}_{mn}^2$, however, the fluctuations in single-particle number no longer cancel as in Ref.~\cite{TrainorReid} and Eq.~(\ref{Eq8}).  Substituting the above summations for ${\cal S}_{mn}$ and ${\cal M}_{mn}$ into $\langle ({\cal S}_{mn} - {\cal M}_{mn})^2 \rangle$ and noting that all cross terms between different events vanish, result in
\bea
\langle ({\cal S}_{mn} - {\cal M}_{mn})^2 \rangle & = & 
\langle \sum_{j=1}^{N_{\rm events}} (\bar{m}^2\nu_j^2 + \bar{n}^2\mu_j^2 + \mu_j^2 \nu_j^2)
 \nonumber \\
 &  & \hspace{-1.0in}  + \sum_{j^{\prime}=1}^{N_{\rm events}} (\bar{m}^2\nu_{j^{\prime}}^2 + \bar{n}^2\mu_{j^{\prime}}^2)
+ \sum_{j^{\prime}=1,{\rm odd}}^{N_{\rm events}}
(\mu_{j^{\prime}}^2 \nu_{j^{\prime}+1}^2 + \mu_{j^{\prime}+1}^2 \nu_{j^{\prime}}^2) \rangle
\nonumber \\
 &  & \hspace{-1.0in} = 2N_{\rm events} \langle \bar{m}^2 \sigma_{\nu}^2 + \bar{n}^2  \sigma_{\mu}^2 + \sigma_{\mu}^2 \sigma_{\nu}^2 \rangle
\stackrel{\rm Poisson}{\longrightarrow}
2N_{\rm events} (\bar{m}^2\bar{n} + \bar{n}^2\bar{m} + \bar{m}\bar{n})
\nonumber
\eea
and the error in ${\cal R}_{mn}$ is
\bea
\sqrt{(\Delta {\cal R}_{mn})^2} & = & \left[ \frac{2}{N_{\rm events}} \left(
\frac{1}{\bar{m}} + \frac{1}{\bar{n}} + \frac{1}{\bar{m}\bar{n}} \right) \right] ^{1/2}.
\nonumber
\eea
Comparing this error to the corresponding statistical error when the same event collection is used for the sibling and mixed-event pairs in Eq.~(\ref{Eq10}) yields the ratio
\bea
\frac{\sqrt{(\Delta {\cal R}_{mn})^2}_{\rm same}}
     {\sqrt{(\Delta {\cal R}_{mn})^2}_{\rm diff}}
 & = & \frac{1}{\sqrt{\bar{m}+\bar{n}+1}}.
\nonumber
\eea
For 200~GeV Au + Au collisions with events selected in the mid-central range $N_{\rm chrg} \in [152,187]$ and for the off-diagonal $(m,n)$ bin on $(y_{t1},y_{t2})$, defined by the ranges ([1.56,1.70],[2.96,3.10]) where $\bar{m}$ = 14.3 and $\bar{n}$ = 3.39, the above ratio of statistical errors is 0.23. The noise reduction of the Reid-Trainor method is more than a factor of 4 in this instance.




\end{document}